\documentclass[12pt]{article}
\usepackage{graphics}
\usepackage{amsthm}
\usepackage{amssymb}
\usepackage{amsmath}
\usepackage{amsfonts}
\usepackage{epic}
\usepackage{hyperref}
\usepackage{epsfig}
\usepackage{bm}
\usepackage{amscd}

\theoremstyle{plain}

\theoremstyle{definition}

\def\be{\begin{equation}}
\def\ee{\end{equation}}
\smallskip
\input{tcilatex}

\begin{document}

\begin{titlepage}
\begin{flushright}
%hep-th/.......\\
\end{flushright}
%%%%%%%%%%%%%%%%%%%%%%%%%%%%%%%%%%%%%%%%%%%%%%%%%%%%%%%%%%%%%%%%%%%%%%%%
\begin{center}
\noindent{{\LARGE{Comment on three-point function in AdS(3)/CFT(2)}}}

\smallskip
\smallskip

\smallskip
\smallskip

\smallskip
\smallskip
\noindent{\large{Gaston Giribet$^1$ and Lorena Nicol\'as$^2$}}

\smallskip
\smallskip

\end{center}
\smallskip
\centerline{$^1$ Departamento de F\'{\i}sica, Universidad de Buenos Aires FCEN - UBA and CONICET}
\centerline{{\it Ciudad Universitaria, Pabell\'on I, 1428. Buenos Aires, Argentina.}}
\smallskip
\smallskip
\smallskip
\centerline{$^2$ Instituto de Astronom\'{\i}a y F\'{\i}sica del Espacio IAFE - CONICET}
\centerline{{\it Ciudad Universitaria, C.C. 67 Suc. 28, 1428, Buenos Aires, Argentina.}}
\smallskip

\smallskip

\smallskip

\begin{abstract} Recently, exact agreement has been found between bulk and
boundary three-point functions in $AdS_3 \times S^3 \times T^4$ with NSNS fluxes. This represents a
non-trivial check of AdS/CFT correspondence beyond the supergravity
approximation as it corresponds to an exact worldsheet computation. When taking a closer look at this computation, one notices that a crucial point for the bulk-boundary agreement to hold is an intriguing mutual cancellation between worldsheet contributions corresponding to the $AdS_3$ and
to the $S^3$ pieces of the geometry, what results in a simple factorized form for the final three-point function. In this note we review this cancellation and clarify some points about the analytic relation
between the $SU(2)$ and the $SL(2,\mathbb{R})$ structure constants. In particular, we dicuss the connection to the Coulomb gas representation. We also make some comments on the four-point function.

\end{abstract}

\end{titlepage}
%%%%%%%%%%%%%%%%%%%%%%%%%%%%%%%%%%%%%%%%%%%%%%%%%%%%%%%%%%%%%%%%%%%%%%%%

%\newpage

%\tableofcontents

\newpage

%%%%%%%%%%%%%%%%%%%%%%%%%%%%%%%%%%%%%%%%%%%%%%%%%%%%%%%%%%%%%%%%%%%%%%

\section{Introduction}

Exact agreement has been observed between boundary and bulk three-point
functions in $AdS_{3}\times S^{3}\times T^{4}$ with NSNS fluxes. In Refs. 
\cite{GK,PD}, Gaberdiel and Kirsch, and Dabholkar and Pakman, computed 
three-point functions of certain chiral primary 
states for
Type IIB\ string theory on $AdS_{3}\times S^{3}\times T^{4}$ in the 
tree-level approximation. And the resulting expressions were compared
with the corresponding correlators in the dual two-dimensional conformal
field theory at the orbifold point. As a result, exact agreement was found
between bulk and boundary observables at large N. In \cite{PS}, Pakman 
and Sever extended the analysis
of this holographic agreement to the case of chiral $\mathcal{N}%
=4$ operators, and the operators of spectral flowed sectors were considered
in Ref. \cite{GPR}. The agreement was also studied from the supergravity
point of view in Ref. \cite{Marika}.

The exact agreement found in \cite{GK,PD,PS} not only represents a highly
non-trivial check of AdS/CFT correspondence beyond the supergravity
approximation, but it can also be seen as evidence that a new
non-renormalization theorem holds for string theory in this background. This
is because the bulk and the boundary computations are performed at different
point of the moduli space. This non-renormalization mechanism was recently
studied in Ref. \cite{Verlinde}.

When going through the worldsheet computation of \cite{GK,PD} one
immediately notices that a crucial point to find agreement between bulk and
boundary observables is the surprising cancellation of all the factors in
the worldsheet three-point functions that mix the momenta of vertex
operators. Since the superstring $\sigma $-model in $AdS_{3}\times
S^{3}\times T^{4}$ with NSNS fluxes corresponds to the $\mathcal{N}=1$
Wess-Zumino-Novikov-Witten (WZNW) on $SL(2,\mathbb{R})\times SU(2)\times
U(1)^{4}$, it turns out that such cancellation gets translated into a
remarkable simplification that happens between $SL(2,\mathbb{R})\ $and $%
SU(2) $ structure constants when both are brought together.

To those who are familiarized with the Minimal Liouville Gravity (or, say
the minimal string theory), the cancellation between $SL(2,\mathbb{R})\ $and 
$SU(2)$ structure constants could seem reminiscent of the simplification
that happens between three-point functions in Liouville Field Theory (LFT)
and the three-point function in the Generalized Minimal Models (GMM). It was
pointed out by Al. Zamolodchikov that, even though the analytic relation
between GMM and LFT might give rise to the idea that GMM observables are
simply an analytic continuation of the LFT quantities for pure imaginary
values of the Liouville parameter $b$, it is not actually the case. It was
shown in \cite{Zamolodchikov} that the GMM structure constants are not the
mere analytic continuation of the LFT ones. In fact, contrary to one's
expectation, GMM structure constants turn out to be, up to a proper
renormalization of the vertex operators, the inverse of LFT structure
constants, in the sense that the product of both quantities yields a
remarkably simple factorized expression like $\sim
\prod\nolimits_{i=1}^{3}f(a_{i})$, where $a_{i}$ are the momenta of the
Liouville vertex operators.

It was noticed in \cite{PD} that the cancellation that takes place between
the $SU(2)$ and the $SL(2,\mathbb{R})$ supersymmetric structure constants
when computing three-string amplitudes in $AdS_{3}\times S^{3}\times T^{4}$
is similar to what happens between GMM and LFT observables. This observation
is correct, but, if not interpreted correctly, it might lead to the wrong
conclusion that $SU(2)$ observables cannot be obtained as the analytic
continuation of the analogous $SL(2,\mathbb{R})$ observables for negative
values of the WZNW level $k$. What we want to point out in this note is
that, unlike what happens in the $\mathcal{N}=1$ supersymmetric WZNW model,
where the product of $SU(2)$ and $SL(2,\mathbb{R})$ three-point functions
yields a simple factorized form as in Minimal Liouville Gravity, the
relation between bosonic $SU(2)$ structure constants and bosonic $SL(2,%
\mathbb{R})$ structure constants is different, and it does admit to be seen
as an analytic continuation in $k$. Such analytic continuation is actually
what one considers in the Coulomb gas approach to the non-rational WZNW
theory.

The paper is organized as follows: In Section 2, we discuss correlation
functions in both $SL(2,\mathbb{R})$ and $SU(2)$ WZNW theory. In Section 3,
we review the calculation of three-point amplitudes of chiral states in $%
AdS_{3}\times S^{3}\times T^{4}$. We focus our attention on the
cancellations that take place between the $AdS_{3}$ and the $S^{3}$
contributions. We discuss the analytic relation between $SL(2,\mathbb{R})$
and $SU(2)$ structure constants. Section 4 contains some concluding remarks.
In particular, we make some comments on the four-point function.

\section{Correlation functions in WZNW theory}

\subsection{$SL(2,\mathbb{R})_{k}$ WZNW correlators from Liouville theory}

The $\mathcal{N}=1$ supersymmetric $SL(2,\mathbb{R})_{\widehat{k}}$ WZNW
model describes the superstring $\sigma $-model on the $AdS_{3}$ piece of
the spacetime, where the relation between the $AdS_{3}$ radius $l$ and the
string length scale $l_{s}$ is given by $\widehat{k}=l^{2}/l_{s}^{2}$, so
that the semiclassical limit corresponds to $\widehat{k}$ large. This
interpretation is consistent with the value of the central charge of the
theory%
\begin{equation*}
c_{sl\text{(2)}}=3+6/\widehat{k}\text{,}
\end{equation*}%
which tends to $3$ when $\widehat{k}$ goes to infinity.

The supersymmetric affine algebra of the WZNW theory is generated by the
supercurrent $\psi ^{a}(z)+\theta J^{a}(z)$, where $a=1,2,3$, $\theta $ is a
Grassman variable, and $\psi ^{a}(z)$ represent three free fermions. The
currents $J^{a}$ generate the affine algebra $\widehat{sl}(2)$ of level $%
\widehat{k}$, which is realized by the following operator product expansion
(OPE)%
\begin{equation*}
J^{a}(z)J^{b}(w)\sim \frac{\eta ^{ab}\widehat{k}/2}{(z-w)^{2}}+\frac{%
i\varepsilon ^{abc}J_{c}(w)}{(z-w)}+...
\end{equation*}%
where $\varepsilon ^{abc}=1$ and $\eta ^{ab}=$ diag$(++-)$, with $%
a,b,c=1,2,3 $. The generators $J^{a}(z)$ can be written as%
\begin{equation*}
J^{a}(z)=j^{a}(z)-\frac{i}{\widehat{k}}\varepsilon ^{abc}\psi _{b}(z)\psi
_{c}(z),
\end{equation*}%
where, in turn, the bosonic currents $j^{a}(z)$ generate $sl(2)_{k}$ of
level $k=\widehat{k}+2$. The OPE between the currents $J^{a}(z)$ and the
fermions $\psi ^{a}(z)$ reads 
\begin{equation*}
J^{a}(z)\psi ^{b}(w)\sim \frac{i\varepsilon ^{abc}\psi _{c}(w)}{(z-w)}%
+...,\quad \quad \psi ^{a}(z)\psi ^{b}(w)\sim \frac{\eta ^{ab}\widehat{k}/2}{%
(z-w)}+...
\end{equation*}

The Sugawara construction yields the stress-tensor%
\begin{equation}
T(z)=\frac{1}{\widehat{k}}\eta _{ab}(J^{a}(z)J^{b}(z)-\psi ^{a}(z)\partial
\psi ^{b}(z))  \label{sugawara}
\end{equation}%
that generates the worldsheet Virasoro algebra.

The vertex operators $\Phi _{j}(x|z)$ representing states of the worldsheet
theory are given by Virasoro primary fields w.r.t. (\ref{sugawara}) and
expand representations of $SL(2,\mathbb{R})$. The index$\ j$ labels such
representation of $SL(2,\mathbb{R})$, while $x$ is an auxiliary complex
variable that allows for the following realization of the algebra%
\begin{equation*}
j^{a}(z)\Phi _{j}(x|w)=-\frac{\mathcal{D}_{x}^{a}\Phi _{j}(x|w)}{(z-w)}+...
\end{equation*}%
with the differential operators%
\begin{equation*}
\mathcal{D}_{x}^{+}=x^{2}\partial _{x}-2jx,\quad \mathcal{D}%
_{x}^{-}=\partial _{x},\quad \mathcal{D}_{x}^{3}=x\partial _{x}-j,
\end{equation*}%
where, as usual, the notation $a=+,-,3$ corresponds to the generators $%
J^{\pm }(z)=J^{1}(z)\pm iJ^{2}(z)$.

The conformal dimension of vertex operators $\Phi _{j}(x|z)$ is given by%
\begin{equation*}
\Delta _{sl\text{(2)}}=-\frac{j(j+1)}{k-2},\quad \text{with}\quad k=\widehat{%
k}-2.
\end{equation*}

Here, we are interested in correlation functions of these vertex operators.
The four-point correlation function in the $SL(2,\mathbb{R})_{k}$ WZNW
theory can be written in terms of the five-point function in LFT as follows 
\cite{T}%
\begin{equation}
\left\langle \prod\nolimits_{i=1}^{4}\Phi _{j_{i}}(x_{i}|z_{i})\right\rangle
_{sl\text{(2)}}=\mathcal{X}_{k}(j_{1},j_{2},j_{3},j_{4}|x,z)\times
\left\langle \prod\nolimits_{i=1}^{5}V_{a_{i}}(z_{i})\right\rangle _{\text{%
LFT}}  \label{karita}
\end{equation}%
where $2a_{1}=-b(j_{1}+j_{2}+j_{2}+j_{4}+1)$, $%
2a_{5>i>1}=-b(j_{1}+2j_{i}-j_{2}-j_{3}-j_{4}-b^{-2}-1)$, $2a_{5}=-b^{-1}$, $%
b^{-2}=k-2$, $z_{1}=z$, $z_{2}=0$, $z_{3}=1$, $z_{4}=\infty $, and on the
r.h.s. also holds that $z_{5}=x$. The correlation function on the r.h.s.
involves five exponential vertex operators of LFT (see (\ref{tornadito})
below). The function $\mathcal{X}_{k}(j_{1},j_{2},j_{3},j_{4}|x,z)$ is given
by%
\begin{equation*}
\mathcal{X}_{k}(j_{1},j_{2},j_{3},j_{4}|x,z)=\frac{%
|z|^{4(a_{1}a_{2}-b^{2}j_{1}j_{2})}|z-1|^{4(a_{1}a_{3}-b^{2}j_{1}j_{3})}}{%
|x|^{2a_{2}b^{-1}}|x-1|^{2a_{3}b^{-1}}|x-z|^{2a_{1}b^{-1}}}%
X_{k}(j_{1},j_{2},j_{3},j_{4})
\end{equation*}%
with\footnote{%
When compareing with \cite{T}, take into account the relations $\Upsilon
_{W}(x)=\Upsilon (-xb)=G_{k}^{-1}(x)b^{-b^{2}x^{2}-(b^{2}+1)x}$.} 
\begin{equation*}
X_{k}(j_{1},j_{2},j_{3},j_{4})=\frac{\pi C_{W}^{2}(b)}{b^{5+4b^{2}}\Upsilon
_{0}^{2}}\frac{\left( \nu (b)\right) ^{s}}{\left( \pi \mu \gamma
(b^{2})b^{4}\right) ^{2j_{1}}}\times 
\end{equation*}%
\begin{equation}
\times \frac{G_{k}\left( 2+\sum_{i=1}^{4}j_{i}\right)
\prod_{n=2}^{4}G_{k}\left( -1-j_{1}-2j_{n}+\sum_{i=2}^{4}j_{i}\right) \gamma
\left( b^{2}\left( j_{1}+2j_{n}-\sum_{i=2}^{4}j_{i}\right) \right) }{\gamma
\left( -b^{2}\sum_{i=1}^{4}j_{i}-2b^{2}\right) \prod_{t=1}^{4}G_{k}(2j_{t}+1)%
},  \label{esaconelAA}
\end{equation}%
where $s=1+\sum_{i=1}^{4}j_{i}$, $\gamma (x)=\Gamma (x)/\Gamma (1-x)$, $\nu
(b)=-b^{2}\gamma (-b^{2})$, and where the special function $G_{k}(x)$ obeys
the functional relations 
\begin{equation}
G_{k}(x)=G_{k}(x-1)\gamma (-b^{2}x),\qquad G_{k}(x)=G_{k}(-1-x-b^{-2}),
\label{Prop2}
\end{equation}%
(see \cite{MO3} and references therein). The overall factor $\frac{\pi
C_{W}^{2}(b)}{b^{2}\Upsilon _{0}^{2}}$ in (\ref{esaconelAA}) is a $b$%
-dependent function (namely, a factor independent of $j_{i}$), and it can be
found in Ref. \cite{T}. The $SL(2,\mathbb{R})_{k}$ structure constants can
be obtained from (\ref{esaconelAA}) in the limit $j_{1}=n=0$.

Eq. (\ref{karita}) relates correlation functions of two different
non-rational theories. It follows from the remarkable observation,
originally due to Fateev and Zamolodchikov \cite{FZ}, that the
Knizhnik-Zamolodchikov equation \cite{KZ} satisfied by the WZNW four-point
function generates a solution to the Belavin-Polyakov-Zamolodchikov equation 
\cite{BPZ} satisfied by the five-point function that involves a degenerate
field of momentum $a_{5}=-1/2b$.

Relation (\ref{karita}) permits to understand several non-trivial properties
of the pole structure of $SL(2,\mathbb{R})_{k}$ WZNW four-point function: In 
\cite{GiribetSimeone} it was shown that the logarithmic dependences in the $%
AdS_{3}$ amplitudes, which can be understood in terms of $AdS_{3}/CFT_{2}$
as in \cite{Rastelli1, Rastelli2}, are ultimately associate to the OPE $%
V_{(b+1/b)/2}(z_{i})V_{-1/2b}(x)$ when $a_{i}=(b+1/b)/2$ for $i=2,3,4$.
Representation (\ref{karita}) is also useful to understand the origin of
poles at the point $z=x$ that are associated to worldsheet instantons \cite%
{MO3}. While from the perspective of WZNW theory such poles are unexpected
as they are located in the middle of the moduli space, in terms of LFT these
are naturally understood as emerging in the coincidence limit of two
operators $V_{-sb/2}(z_{1})V_{-1/2b}(x)$.

The normalization $X_{k}(j_{1},j_{2},j_{3},j_{4})$ in (\ref{karita}) is
compatible with crossing symmetry of WZNW theory \cite{T}. It can be also
shown that $X_{k}(j_{1},j_{2},j_{3},j_{4})$ leads to a nice realization of
the Hamiltonian reduction, which here corresponds to the limit $x\rightarrow
z$ \cite{Furlan1, Furlan2}. In this limit, and considering the OPE%
\begin{eqnarray*}
V_{a_{i}}(z_{i})V_{-1/2b}(x) &=&\left| x-z\right| ^{2\xi
_{-}}V_{-1/2b+a_{i}}(z_{i})+ \\
&&+(\pi \mu \gamma (b^{2}))^{b^{-2}}\frac{\gamma (2a_{i}b^{-1}-1-b^{-2})}{%
b^{4}\gamma (2a_{i}b^{-1})}\left| x-z\right| ^{2\xi
_{+}}V_{-1/2b-a_{i}}(z_{i}),
\end{eqnarray*}%
with $\xi _{\pm }=(\Delta _{a_{i}\pm 1/2b}-\Delta _{1/2b}-\Delta _{a_{i}})$
and $\Delta _{a}=a(b+b^{-1}-a)$, one finds%
\begin{equation}
\left\langle \prod\nolimits_{i=1}^{4}\Phi _{j_{i}}(x_{i}|z_{i})\right\rangle
_{sl\text{(2)}}\sim \prod\nolimits_{i=1}^{4}\gamma (1+b^{2}(2j_{i}+1))\times
\left\langle \prod\nolimits_{i=1}^{4}V_{-bj_{i}}(z_{i})\right\rangle _{\text{%
LFT}}+...,  \label{HR}
\end{equation}%
where the symbol $\sim $ stands for a $b$-dependent factor and a singular
factor $\left| x-z\right| ^{-2(1+b^{-2})}$, while the ellipses stand
for subleading contribution, provided the Seiberg bound $a_{i}>(b+b^{-1})/2$
is obeyed. Notice that factors $\gamma (1+b^{2}(2j_{i}+1))$ in (\ref{HR})
can be absorbed in the normalization of Liouville vertices. Expression (\ref%
{HR}) can be proven by using formulas (1.28)-(1.29) of \cite{FL}, together
with the kind of tricks used in the Appendix B of \cite{GiribetSimeone}.

The Liouville correlation functions in (\ref{esaconelAA}) are defined by%
\begin{equation}
\left\langle \prod\nolimits_{i=1}^{5}V_{a_{i}}(z_{i})\right\rangle _{\text{%
LFT}}=\int D\varphi e^{-S_{L}[\varphi ;\mu ]}\prod\nolimits_{i=1}^{5}e^{%
\sqrt{2}a_{i}\varphi (z_{i})},  \label{tornadito}
\end{equation}%
with 
\begin{equation}
S_{L}[\varphi ;\mu ]=\frac{1}{4\pi }\int d^{2}z\left( \partial \varphi 
\overline{\partial }\varphi +(b+b^{-1})R\varphi /2\sqrt{2}+4\pi \mu e^{\sqrt{%
2}b\varphi }\right) ,
\end{equation}%
where $R$ is the scalar curvature of the worldsheet and $\mu $ is a real
parameter \cite{N}. By integrating out the zero-mode of $\varphi $, (\ref%
{tornadito}) can be expanded as%
\begin{eqnarray}
\left\langle \prod\nolimits_{i=1}^{5}V_{a_{i}}(z_{i})\right\rangle _{\text{%
LFT}} &=&\Gamma (-n)b^{-1}\mu ^{n}\delta
(nb+\sum\nolimits_{i=1}^{5}a_{i}-b-b^{-1})\times  \notag \\
&&\times \int D\varphi e^{-S_{L}[\varphi ;\mu =0]}\prod\nolimits_{i=1}^{5}e^{%
\sqrt{2}a_{i}\varphi (z_{i})}\prod\nolimits_{r=1}^{n}e^{\sqrt{2}b\varphi
(w_{r})},  \label{SSSS}
\end{eqnarray}%
where now the path integral is understood as not including the zero-mode 
\cite{GL}.

It is important to notice that expression above admits an integral
representation of the form%
\begin{equation*}
\left\langle \prod\nolimits_{i=1}^{4}\Phi _{j_{i}}(x_{i}|z_{i})\right\rangle
_{sl\text{(2)}%
}=X_{k}(j_{1},j_{2},j_{3},j_{4})|z|^{-4b^{2}j_{1}j_{2}}|1-z|^{-4b^{2}j_{1}j_{3}}\Gamma (-n)b^{-1}\mu ^{n}\times
\end{equation*}%
\begin{equation}
\times \int
\prod\nolimits_{r=1}^{n}d^{2}w_{r}\prod%
\nolimits_{r=1}^{n}|w_{r}|^{-4a_{2}b}|w_{r}-1|^{-4a_{3}b}|w_{r}-z|^{-4a_{1}b}|w_{r}-x|^{2}\prod\nolimits_{r<t}^{n}|w_{r}-w_{t}|^{-4b^{2}},
\label{ponelenombre}
\end{equation}%
where 
\begin{equation}
n=b+b^{-1}(1-\sum\nolimits_{i=1}^{5}\alpha _{i})=2j_{1}.  \label{nuevaq}
\end{equation}

As mentioned, for the particular case $j_{1}=0$ we would obtain the
structure constants $C_{sl\text{(2)}}(j_{2},j_{3},j_{4})\sim
X_{k}(0,j_{2},j_{3},j_{4})$. Replacing $j_{1}=0$ in the equation above it
yields 
\begin{eqnarray}
\mathcal{X}_{k}(0,j_{2},j_{3},j_{4}) &=&-\frac{\gamma (-b^{2})}{2\pi ^{2}}%
(\nu (b))^{j_{2}+j_{3}+j_{4}+1}\frac{G_{k}(1+j_{2}+j_{3}+j_{4})}{G_{k}(-1)}%
\times  \notag \\
&&\times \frac{%
G_{k}(-j_{2}+j_{3}+j_{4})G_{k}(j_{2}-j_{3}+j_{4})G_{k}(j_{2}+j_{3}-j_{4})}{%
G_{k}(2j_{2}+1)G_{k}(2j_{3}+1)G_{k}(2j_{4}+1)}  \label{esaconelA}
\end{eqnarray}%
where the overall factor $\frac{C_{W}^{2}(b)G_{k}(-1)}{\Upsilon
_{0}^{2}G_{k}(1)}$ has been replaced by $\frac{b^{1+4b^{2}}}{2\pi ^{3}}%
\gamma (1-b^{2})$, taking into account that in the limit $j_{1}\rightarrow 0$
one finds $\frac{G_{k}(-1)}{G_{k}(1)}\Gamma (-n)=\frac{b^{2}}{2}\gamma
(1+b^{2})$.

Now, let us move on and consider the four-point function in the $SU(2)_{k}$
model.

\subsection{$SU(2)_{k}$ WZNW correlators from Minimal Models}

The $\mathcal{N}=1$ supersymmetric $SU(2)_{\widehat{k}}$ WZNW theory has
central charge%
\begin{equation*}
c_{su\text{(2)}}=3-6/\widehat{k}\text{.}
\end{equation*}

The affine symmetry is generated by the current algebra $\widehat{su}(2)_{%
\widehat{k}}$, realized by the OPE%
\begin{equation*}
K^{a}(z)K^{b}(w)\sim \frac{\delta ^{ab}\widehat{k}/2}{(z-w)^{2}}+\frac{%
i\varepsilon ^{abc}K_{c}(w)}{(z-w)}+...
\end{equation*}%
where $\varepsilon ^{abc}=1$ and now $\delta ^{ab}=$diag$(+++)$, with $%
a,b,c=1,2,3$. It also holds that%
\begin{equation*}
K^{a}(z)\chi ^{b}(w)\sim \frac{i\varepsilon ^{abc}\chi _{c}(w)}{(z-w)}%
+...,\quad \quad \chi ^{a}(z)\chi ^{b}(w)\sim \frac{\delta ^{ab}\widehat{k}/2%
}{(z-w)}+...
\end{equation*}

As in the case of $SL(2,\mathbb{R})_{k}$, the generators can be written as%
\begin{equation*}
K^{a}(z)=k^{a}(z)-\frac{i}{\widehat{k}}\varepsilon ^{abc}\chi _{b}(z)\chi
_{c}(z),
\end{equation*}%
where the bosonic currents $k^{a}(z)$ generate the algebra $\widehat{su}%
(2)_{k^{\prime }}$ of level $k^{\prime }=\widehat{k}-2$, and $\chi ^{a}(z)$
represent three free fermions.

The vertex operators $\Psi _{j^{\prime }}(y|z)$ are Virasoro primaries of
conformal dimension%
\begin{equation*}
\Delta _{su\text{(2)}}=\frac{j(j+1)}{k^{\prime }+2},\quad \text{with}\quad
k^{\prime }=\widehat{k}-2.
\end{equation*}%
where$\ j^{\prime }$ now labels representation of $SU(2)$, and where, again, 
$y$ is an auxiliary complex variable such that%
\begin{equation*}
k^{a}(z)\Psi _{j^{\prime }}(y|w)=-\frac{\mathcal{K}_{y}^{a}\Psi _{j^{\prime
}}(y|w)}{(z-w)}+...
\end{equation*}%
with%
\begin{equation*}
\mathcal{K}_{y}^{+}=y^{2}\partial _{y}-2j^{\prime }y,\quad \mathcal{K}%
_{y}^{-}=-\partial _{y},\quad \mathcal{K}_{y}^{3}=y\partial _{y}-j^{\prime },
\end{equation*}%
and with $K^{\pm }(z)=K^{1}(z)\pm iK^{2}(z)$.

Four-point correlation function in the $SU(2)_{k^{\prime }}$ WZNW theory can
be written in terms of the five-point function in GMM as follows%
\begin{equation}
\left\langle \prod\nolimits_{i=1}^{4}\Psi _{j_{i}^{\prime
}}(y_{i}|z_{i})\right\rangle _{su\text{(2)}}=\mathcal{Y}_{k^{\prime
}}(j_{1}^{\prime },j_{2}^{\prime },j_{3}^{\prime },j_{4}^{\prime
}|y,z)\times \left\langle \prod\nolimits_{i=1}^{5}W_{\alpha
_{i}}(z_{i})\right\rangle _{\text{GMM}}  \label{FZ}
\end{equation}%
where $2\alpha _{1}=\beta (j_{1}+j_{2}+j_{2}+j_{4}+1)$, $2\alpha
_{5>i>1}=\beta (j_{1}+2j_{i}-j_{2}-j_{3}-j_{4}+k^{\prime }+1)$, $2\alpha
_{5}=\beta ^{-1}$, $\beta ^{-2}=k^{\prime }+2$, $z_{1}=z$, $z_{2}=0$, $%
z_{3}=1$, $z_{4}=\infty $, and on the r.h.s. also holds that $z_{5}=y$. Eq. (%
\ref{FZ}) is the $SU(2)$ analogue of (\ref{karita}). Function $\mathcal{Y}%
_{k}(j_{1},j_{2},j_{3},j_{4}|y,z)$ is given by 
\begin{equation*}
\mathcal{Y}_{k^{\prime }}(j_{1}^{\prime },j_{2}^{\prime },j_{3}^{\prime
},j_{4}^{\prime }|y,z)=\frac{|z|^{4(\beta ^{2}j\prime _{1}j\prime
_{2}-\alpha _{1}\alpha _{2})}|z-1|^{4(\beta ^{2}j\prime _{1}j\prime
_{3}-\alpha _{1}\alpha _{3})}}{|y|^{-2\alpha _{2}\beta ^{-1}}|y-1|^{-2\alpha
_{3}\beta ^{-1}}|y-z|^{-2\alpha _{1}\beta ^{-1}}}Y_{k^{\prime
}}(j_{1}^{\prime },j_{2}^{\prime },j_{3}^{\prime },j_{4}^{\prime })
\end{equation*}%
with%
\begin{eqnarray}
Y_{k^{\prime }}(j_{1}^{\prime },j_{2}^{\prime },j_{3}^{\prime
},j_{4}^{\prime }) &=&\left( \gamma (\beta ^{2})\right) ^{2j_{1}^{\prime
}+1}P_{k^{\prime }}\left( \sum\nolimits_{a=1}^{4}j_{a}^{\prime }+1\right)
\prod_{i=1}^{4}\frac{\sqrt{\gamma (1-\beta ^{2}(2j_{i}^{\prime }+1))}}{%
P_{k^{\prime }}(2j_{i}^{\prime })}\times  \notag \\
&&\times \prod_{n=2}^{4}P_{k^{\prime }}\left(
\sum\nolimits_{l=2}^{4}j_{l}^{\prime }-2j_{n}^{\prime }-j_{1}^{\prime
}\right)  \label{Otraduda}
\end{eqnarray}%
where,%
\begin{equation*}
P_{k^{\prime }}(x)=\prod_{n=1}^{x}\gamma \left( n\beta ^{2}\right) ,\qquad
x\geq 1,
\end{equation*}%
while $P_{k^{\prime }}(0)=1$. Normalization factor (\ref{Otraduda}) is
consistent with the fusion rules of the algebra \cite{Andreev}.

Expression (\ref{FZ}) above also admits an integral representation \cite%
{FZ,DF,DF2}; namely%
\begin{equation*}
\left\langle \prod\nolimits_{i=1}^{4}\Psi _{j_{i}^{\prime
}}(y_{i}|z_{i})\right\rangle _{su\text{(2)}}=Y_{k^{\prime }}(j_{1}^{\prime
},j_{2}^{\prime },j_{3}^{\prime },j_{4}^{\prime })|z|^{4\beta ^{2}j\prime
_{1}j\prime _{2}}|1-z|^{4\beta ^{2}j\prime _{1}j\prime _{3}}\times
\end{equation*}%
\begin{equation*}
\times \int \prod\nolimits_{r=1}^{2j\prime
_{1}}d^{2}t_{r}\prod\nolimits_{r=1}^{2j\prime _{1}}|t_{r}|^{-4\alpha
_{2}\beta }|t_{r}-1|^{-4\alpha _{3}\beta }|t_{r}-z|^{-4\alpha _{1}\beta
}|t_{r}-y|^{2}\prod\nolimits_{r<l}|t_{r}-t_{l}|^{4\beta ^{2}}.
\end{equation*}%
This completes the parallelism with the formula (\ref{ponelenombre}) for $%
SL(2,\mathbb{R})$. Now, we are ready to discuss string amplitudes in $%
AdS_{3}\times S^{3}$ in terms of correlation functions of the $SL(2,\mathbb{R%
})\times SU(2)$ theory.

\section{String amplitudes in $AdS_{3}\times S^{3}$}

\subsection{AdS$_{3}$/CFT$_{2}$ correspondence and three-point function}

According to the AdS$_{3}$/CFT$_{2}$ correspondence, correlation functions
of dimension-$h$ operators in the boundary CFT correspond to string
amplitudes on $AdS_{3}$; namely

\begin{equation}
\prod\nolimits_{i=3}^{N}\int d^{2}z_{i}\left\langle
\prod\nolimits_{i=1}^{N}\Phi _{j_{i}}(x_{i}|z_{i})\right\rangle _{\text{%
worldsheet}}\times ...=\left\langle
\prod\nolimits_{i=1}^{N}O_{h_{i}}(x_{i})\right\rangle _{\text{boundary}}
\label{adscft}
\end{equation}%
where the ellipses reflect the contribution of the internal space\footnote{%
More precisely, the complete prescription for AdS$_{3}$/CFT$_{2}$ would also
include contributions coming from disconnected worldsheet diagrams \cite%
{Ooguri}.}.

The indices $j_{i}$, which label the representations of $SL(2,\mathbb{R})$,
are related to the conformal dimension $h_{i}$ of vertex operators in the
dual theory by the simple relation%
\begin{equation}
h_{i}=-j_{i}.  \label{hj}
\end{equation}%
This can be seen, for instance, by looking at the $x$-dependence of
three-point functions in the $SL(2,\mathbb{R})_{k}$ WZNW model, which goes
like 
\begin{equation*}
\left\langle \Phi _{j_{1}}(x_{1}|0)\Phi _{j_{2}}(x_{2}|1)\Phi
_{j_{3}}(x_{3}|\infty )\right\rangle _{sl\text{(2)}}=|x_{12}|^{2\left(
j_{1}+j_{2}-j_{3}\right) }|x_{23}|^{2\left( j_{2}+j_{3}-j_{1}\right)
}|x_{13}|^{2\left( j_{3}+j_{1}-j_{2}\right) }C_{sl\text{(2)}%
}(j_{1},j_{2},j_{3})\text{ },
\end{equation*}%
where $|x_{ij}|=|x_{i}-x_{j}|$. From this we also observe that auxiliary
complex variables $x_{i}$ acquire now a physical meaning, as these are
interpreted as the coordinates of the boundary, where the dual CFT$_{2}$ is
defined on.

Unitarity of the worldsheet theory in $AdS_{3}$ also demands the bound%
\begin{equation}
1-k<2j<-1,\qquad k>2\text{,}  \label{unitaritybound}
\end{equation}%
as well as the introduction of the specral flowed sectors of the $\widehat{sl%
}(2)_{\widehat{k}}$ algebra, which represent winding strings states in $%
AdS_{3}$; see \cite{MO3} and references therein.

The boundary two-dimensional conformal field theory that is dual to the type
IIB string theory in $AdS_{3}\times S^{3}\times T^{4}$ is some deformation
of the symmetric product orbifold Sym$^{\text{N}}(\widetilde{T}^{4})$ of N
copies of $\widetilde{T}^{4}$ \cite{SeibergWitten}, where $\widetilde{T}^{4}$
is closely related to $T^{4}$. This three-dimensional example of holographic
correspondence is motivated by the near horizon limit of the $D1$/$D5$
system, where the geometry $AdS_{3}\times S^{3}\times T^{4}$ is seen to
emerge. In the S-dual picture, this configuration corresponds to the setting
of $Q_{5}=\widehat{k}$ NS5-branes and $Q_{1}$ fundamental strings, where the
number of copies of $\widetilde{T}^{4}$ is given by N$=Q_{1}Q_{5}$. The
six-dimensional string coupling constant is given by $g_{6}^{2}=Q_{5}/Q_{1}$%
, and thus the string perturbative theory is reliable in the large $Q_{1}$
limit, or N$=Q_{5}Q_{1}>>Q_{5}$. In this limit, string states in the bulk
are mapped to twisted states in Sym$^{\text{N}}(\widetilde{T}^{4})$ that are
associated to conjugancy classes with a single non-trivial cycle of length $%
n $. The relation between $n$ and the worldsheet momentum is \cite%
{Argurio,GPR}

\begin{equation*}
n=2h-1+\widehat{k}\omega ,\quad \quad 2h=2,3,4,...k,\quad \quad \omega
=0,1,2,...
\end{equation*}%
where $h$ is associated to index $j$ of the representations of $SL(2,\mathbb{%
R})$ by (\ref{hj}), while $\omega $ labels the spectral flow sector of $SL(2,%
\mathbb{R})$ the representation belongs to. Here we will consider the sector 
$\omega =0$.

We are interested in worldsheet vertex operators that represent chiral
string states in $AdS_{3}\times S^{3}\times T^{4}$. As an example, let us
consider the worldsheet vertex operators of the form%
\begin{equation}
\mathcal{O}_{j}(x|z)=\psi (x|z)\times \Phi _{j}(x|z)\times \Psi _{-1-j}(x|z),
\label{V2}
\end{equation}%
where the fermionic contributions takes the form $\psi (x|z)=-\psi
^{+}(z)+2x\psi ^{3}(z)-x^{2}\psi ^{-}(z)$. This is a worldsheet vertex
operator associated to chiral string states of the NS sector, written in the
picture $-1$. In order to compute a three-point function we also need the
expression for such a state in the picture $0$. This is obtained by reading
off the coefficient of the single pole of the OPE between the worldsheet
supercurrent G$(z)$ and the vertex $\mathcal{O}_{j}(x|z)$. It yields the
following form for the vertex in the picture $0$%
\begin{equation}
\widetilde{\mathcal{O}}_{j}(x|z)=(J(x|z)+\frac{2}{\widehat{k}}\psi (x|z)\psi
_{a}(z)\mathcal{D}_{x}^{a}+\frac{2}{\widehat{k}}\psi (x|z)\chi _{a}(z)%
\mathcal{K}_{x}^{a})\mathcal{O}_{j}(x|z)  \label{V3}
\end{equation}%
where $J(x|z)=-J^{+}(z)+2xJ^{3}(z)-x^{2}J^{-}(z)$. From this, we see that
the computation of the three-point amplitude $\left\langle \mathcal{O}%
_{j_{1}}(x_{1}|z_{1})\mathcal{O}_{j_{2}}(x_{2}|z_{2})\widetilde{\mathcal{O}}%
_{j_{3}}(x_{3}|z_{3})\right\rangle $ also requires to compute correlators of
the form $i\varepsilon _{\text{ }cd}^{f}\left\langle \psi ^{a}(z_{1})\psi
^{b}(z_{2})\psi ^{c}(z_{3})\psi ^{d}(z_{3})\right\rangle $ as well as
correlators that involve the insertion of current operator $J^{a}(z_{3})$.

According to (\ref{adscft}), worldsheet operators $\mathcal{O}%
_{j_{i}}(x_{i}|z_{i})$ are associated to operators $O_{h_{i}}(x_{i})$ in the
boundary CFT. The relation between $SL(2,\mathbb{R})$ spin $j$ and $SU(2)$
spin $j^{\prime }$ in (\ref{V2}) is such that the bosonic contribution to
the conformal dimension corresponding to the $AdS_{3}\times S^{3}$ piece%
\footnote{%
Also notice that the relation between $k^{\prime }$ and $k$ is such that the
total central charge of the worldsheet theory saturates $c=\frac{3k}{k-2}-%
\frac{3k^{\prime }}{k^{\prime }+2}+9=3+\frac{6}{\widehat{k}}+3-\frac{6}{%
\widehat{k}}+4+5=15$, where the contribution of the $T^{4}$ factor and of
the free fermions were included.} of the $\sigma $-model gives $\Delta _{sl%
\text{(2)}}+\Delta _{su\text{(2)}}=0$.

In \cite{GK,PD} three-point functions of chiral operators (\ref{V2}) were
shown to agree with three-point functions in the symmetric product at the
orbifold point. In Refs. \cite{PD,GPR} the computation of all the cases is
discussed in detail. The worldsheet three-point function of chiral operators
(\ref{V2}) (with one of them written in the picture $0$, as in (\ref{V3}))
takes the form%
\begin{equation}
\left\langle \mathcal{O}_{j_{1}}(0|0)\mathcal{O}_{j_{2}}(1|1)\widetilde{%
\mathcal{O}}_{j_{3}}(\infty |\infty )\right\rangle _{\text{worldsheet}}=%
\widehat{k}^{2}|2-h_{1}-h_{2}-h_{3}|^{2}C(-h_{2},-h_{3},-h_{4})  \label{FFFG}
\end{equation}%
where $C(j_{2},j_{3},j_{4})$ is given by the product of $SL(2,\mathbb{R})$
and $SU(2)$ structure constants, namely $C(j_{2},j_{3},j_{4})=C_{sl\text{(2)}%
}(j_{2},j_{3},j_{4})C_{su\text{(2)}}(-1-j_{2},-1-j_{3},-1-j_{4})$; that is, 
\begin{equation}
C(j_{2},j_{3},j_{4})\sim
X_{k}(0,j_{2},j_{3},j_{4})Y_{k-4}(0,-1-j_{2},-1-j_{3},-1-j_{4}).
\end{equation}%
When all the pieces are brought together, and after some manipulation,
expression (\ref{FFFG}) can be seen to agree with the three-point functions
of the boundary theory \cite{LM,LM2}. This agreement exhibited by bulk and
boundary observables is exact, and several steps through the computations
combine in such a subtle form that no doubt remains about this is a highly
non-trivial check of AdS/CFT conjecture. The role played by the
picture-changing operator in the three-point function and by the precise
normalization of the two-point functions are crucial ingredients in the
calculation. Nevertheless, the most striking feature in the calculation is,
so far, the fact that all the dependences that mix the momenta $j_{i}$ in
the three-point function $C_{sl\text{(2)}}(j_{2},j_{3},j_{4})$ cancel out
against analogous dependences coming from $C_{su\text{(2)}%
}(-1-j_{2},-1-j_{3},-1-j_{4})$. In the next subsections we will review these
cancellations and, more interestingly, we will explain why this fact does
not confront the analytic relation that exists between $SL(2,\mathbb{R})$
and $SU(2)$ structure constants.

\subsection{Cancellations in the supersymmetric three-point function}

Let us consider three-point amplitudes of chiral states in type IIB string
theory in $AdS_{3}\times S^{3}\times T^{4}$. The bosonic part corresponding
to the six dimensional piece $AdS_{3}\times S^{3}$ is the non-trivial
contribution here. It is given by correlation functions of vertex operators $%
\Phi _{j}(x|z)\Psi _{-1-j}(x|z)$, which are the product of correlation
functions in the $SL(2,\mathbb{R})_{\widehat{k}+2}$ model and correlation
functions in the $SU(2)_{\widehat{k}-2}$ model, provided the relations $%
\widehat{k}=k-2=k^{\prime }+2$ and $j_{i}=-1-j_{i}^{\prime }$.

Three-point function in $SL(2,\mathbb{R})_{k}$ WZNW model is obtained from (%
\ref{esaconelAA}) by taking the limit $j_{1}\rightarrow 0$. This yields%
\begin{equation}
X_{k}(0,j_{2},j_{3},j_{4})=\frac{\left( \nu (b)\right) ^{j_{2}+j_{3}+j_{4}+1}%
}{2\pi ^{2}\gamma (b^{2})b^{4}}\frac{G_{k}(1+j_{2}+j_{3}+j_{4})}{G_{k}(-1)}%
\prod\nolimits_{i=2}^{4}\frac{G_{k}(j_{2}+j_{3}+j_{4}-2j_{i})}{%
G_{k}(2j_{i}+1)},  \label{gato}
\end{equation}%
as we wrote in (\ref{esaconelA}). On the other hand, knowing that $\beta =b$%
, and being aware that if $x$ is a positive integer then the following
identity holds

\begin{equation*}
P_{k^{\prime }}(x)=\prod_{n=1}^{x}\gamma \left( n\beta ^{2}\right) =\frac{%
G_{k}(-1)}{G_{k}(-1-x)},\qquad x\geq 1,
\end{equation*}%
we obtain the explicit form for the three-point function in the $%
SU(2)_{k^{\prime }}$ WZNW model%
\begin{equation}
Y_{k^{\prime }}(0,-1-j_{2},-1-j_{3},-1-j_{4})=\frac{\sqrt{\gamma (b^{2})}%
G_{k}(-1)}{G_{k}(1+j_{2}+j_{3}+j_{4})}\prod\nolimits_{i=2}^{4}\frac{\sqrt{%
\gamma \left( 1+b^{2}(2j_{i}+1)\right) }G_{k}(2j_{i}+1)}{%
G_{k}(j_{2}+j_{3}+j_{4}-2j_{i})}.  \label{liebre}
\end{equation}%
Rewriting this in a more convenient way and putting both (\ref{gato}) and (%
\ref{liebre}) together, we find

\begin{equation}
X_{k}(0,j_{2},j_{3},j_{4})Y_{k^{\prime }}(0,-1-j_{2},-1-j_{3},-1-j_{4})=%
\frac{1}{2\sqrt{\pi }}\prod\nolimits_{i=2}^{4}\sqrt{B(j_{i})},  \label{RB}
\end{equation}%
where $B(j)$ is given by the $SL(2,\mathbb{R})_{k}$ reflection coefficient, 
\begin{equation*}
\left\langle \Phi _{j_{1}}(x_{1}|0)\Phi _{j_{1}}(x_{2}|1)\right\rangle
=|x_{12}|^{4j_{1}}B(j_{1}),\ \text{with}\ \ B(j)=\left( \nu (b)\right)
^{2j+1}\frac{1}{\pi b^{2}}\gamma (1+(2j_{i}+1)b^{2}).
\end{equation*}

From (\ref{RB}) we observe that the contributions that mixed the momenta $%
j_{i}$ have disappeared. Functions $G_{k}$ coming from both $SL(2,\mathbb{R}%
)_{k}$ and $SU(2)_{k^{\prime }}$ factors cancel against each other, yielding
a rather simplified factorized form. Therefore, we have reproduced the
computation of \cite{GK} and \cite{PD} in a very succinct way, showing that
the three-point function of chiral states in $AdS_{3}\times S^{3}\times
T^{4} $ simplifies in such a way that the dependence of the momenta appear
completely factorized.

Nevertheless, it is worth mentioning that the way we obtained (\ref{RB}) is
not particularly useful, as it is almost the same that working out the
expressions for both $SU(2)_{k^{\prime }}$ and $SL(2,\mathbb{R})_{k}$
structure constants directly, as in \cite{GK,PD,notice}. However, what does
represent an actual advantage is looking at the four-point function in terms
of this minimal gravity representation (see (\ref{esaconelB}) below).

\subsection{Two relations between $SL(2,\mathbb{R})_{k}$ and $SU(2)_{k}$
structure constants}

We have just seen that in the supersymmetric theory, the three-point
function of the $SL(2,\mathbb{R})_{k}$ model and that of $SU(2)_{k^{\prime
}} $ model are related by

\begin{equation}
X_{k}(0,j_{2},j_{3},j_{4})\sim \frac{\prod\nolimits_{i=2}^{4}\sqrt{B(j_{i})}%
}{Y_{k^{\prime }}(0,-1-j_{2},-1-j_{3},-1-j_{4})},  \label{RE}
\end{equation}%
with $k^{\prime }+2=k-2$. That is, all the contributions that mix the
momenta $j_{i}$ in (\ref{esaconelAA}) and (\ref{Otraduda}) disappeared in (%
\ref{RB}). As mentioned, this striking simplification yielding the
factorized form (\ref{RE}) is crucial to find agreement between bulk and
boundary observables.

Expression (\ref{RE}) is due to the relations $j_{i}^{\prime }=-1-j_{i}$ and 
$k-2=k^{\prime }+2$. Roughly speaking, (\ref{RE}) expresses that
supersymmetric $SL(2,\mathbb{R})_{k}$ structure constants are the \textit{%
inverse} of supersymmetric $SU(2)_{k^{\prime }}$ ones, provided the precise
relations between $j_{i}^{\prime }$ and $j_{i}$. In turn, (\ref{RE}) is
analog to the relation between three-point functions in GMM and three-point
functions in LFT \cite{Zamolodchikov}.

Then, a natural question arises: Doesn't this inverse proportionality
relation confront the fact that one can analytically continue the
expressions from $SU(2)_{k}$ to gets its non-compact analog $SL(2,\mathbb{R}%
)_{k}$ (instead of its inverse)? That is, naively one would expect to find
the expression for $SL(2,\mathbb{R})_{k}$ correlators by reversing the sign
of $k$ in the formulas for $SU(2)_{k}$ and performing some analytic
extension; getting something like%
\begin{equation}
X_{k}(0,j_{2},j_{3},j_{4})\sim
Y_{-k}(0,j_{2},j_{3},j_{4})\prod\nolimits_{i=2}^{4}\sqrt{B(j_{i})}.
\label{ER}
\end{equation}

We will see in the next subsection that this is actually the case. That is,
one can analytically continue the expressions and prove a relation like (\ref%
{ER}). We emphasize that this is not in contradiction with relation (\ref{RE}%
) as it is commonly asserted.

\subsection{Analytic continuation in $k$ and the bosonic three-point function%
}

Three-point functions in Minimal Models coupled to Liouville Gravity were
computed by Dotsenko in Ref. \cite{Dotsenko}. When going through the
computation of these correlators, which is based on the Coulomb gas
approach, one needs to make sense of expressions typically given by formal
products of the form%
\begin{equation}
\prod_{n=1}^{x}f(n)  \label{monster}
\end{equation}%
for negative values of the upper index $x$. We will see below that similar
expressions appear when trying to extend the $SU(2)_{k}$ structure constants
for negative values of $j_{i}^{\prime }$ and $k$. In order to propose a
reasonable extension for products like (\ref{monster}) when $x<0$, one can
start by noticing that for positive $x$ it holds 
\begin{equation}
\Pi _{f}(x)=\prod_{n=1}^{x}f(n)=\frac{\prod_{n=1}^{\infty }f(n)}{%
\prod_{n=x+1}^{\infty }f(n)}=\frac{\prod_{n=1}^{\infty }f(n)}{%
\prod_{n=1}^{\infty }f(n+x)}.
\end{equation}%
After that, in a quite natural way, the following extension for the $\Pi
_{f}(x)$ function with negative argument is proposed \cite{Dotsenko},%
\begin{equation}
\Pi _{f}(-x)=\prod_{n=0}^{x-1}f^{-1}(-n).  \label{Prescription}
\end{equation}

Now, consider this analytic extensions for the products $P_{k}(x)$ standing
in (\ref{Otraduda}). It yields%
\begin{equation}
\prod_{n=1}^{-|l|}\gamma (nb^{2})=b^{4(|l|-1)}\frac{\gamma (|l|)}{\gamma
(|l|b^{2})}\prod_{n=1}^{+|l|}\gamma (nb^{2})
\end{equation}%
being $l$ an integer and where we used $\gamma (x)\gamma (1-x)=1$, $\gamma
(1+x)=-x^{2}\gamma (x)$. This permits to make sense of the following
expression 
\begin{equation}
\prod_{n=-|l|}^{+|l|}\gamma (nb^{2})=\gamma (-|l|)b^{-4|l|-2},
\end{equation}%
which will be rederived later in an alternative way.

Now, let us use (\ref{Prescription}) to show how the $SL(2,\mathbb{R})_{k}$
structure constants can be obtained by analytic extension of the $%
SU(2)_{k^{\prime }}$ quantities, provided the relation $k^{\prime }=-k$.
Although it might seem we have already shown this, it is worth noticing that
what we showed before is something slightly different: We showed that, if $%
k^{\prime }+2=k-2$ and $j_{i}^{\prime }=-1-j_{i}$, then the $SL(2,\mathbb{R}%
)_{k}$ structure constants are inversely proportional to $SU(2)_{k^{\prime
}} $ structure constants.

To derive $SL(2,\mathbb{R})_{k}$ structure constants from (\ref{Otraduda})
we assume $k=-k^{\prime }$, and then write%
\begin{equation*}
P_{k^{\prime }}(x)=\prod_{n=1}^{x}\gamma \left( n\beta ^{2}\right)
=\prod_{n=1}^{x}\gamma \left( -nb^{2}\right) =\prod_{n=1}^{x}\gamma
^{-1}\left( 1+nb^{2}\right)
\end{equation*}%
since now $\beta ^{2}=\frac{1}{k^{\prime }+2}=-b^{2}=-\frac{1}{k-2}$
(instead of $\beta ^{2}=+b^{2}$ as before). According to (\ref{Prescription}%
), for $x<0$ we have%
\begin{equation}
P_{k^{\prime }}(x)=\frac{\gamma (|x|b^{2})}{\Gamma (0)}\left(
\prod\nolimits_{n=1}^{|x|}\gamma \left( nb^{2}\right) \right) ^{-1}=\frac{%
G_{k}(-1-|x|)}{\Gamma (0)G_{k}(-1)}\gamma \left( |x|b^{2}\right) ,\qquad x<0.
\label{kopada}
\end{equation}

That is, if $k=-k^{\prime }$ and $x<0$ we get $P_{-k}(x)\sim
G_{k}(x-1)\gamma \left( -xb^{2}\right) /G_{k}(-1)$, while if $k-2=k^{\prime
}+2$ and $x>0$ we get something different like $P_{-k}(x)\sim
G_{k}(-1)/G_{k}(-1-x)$. Using expression (\ref{kopada}) and $%
G_{k}(x)=G_{k}(-1+x)\gamma \left( -xb^{2}\right) $ one finds\footnote{%
Here, we have omitted a divergent $\Gamma (0)$ factor; see discussion below.}%
\begin{eqnarray}
Y_{-k}(0,j_{2},j_{3},j_{4}) &=&\frac{\sqrt{\gamma \left( -b^{2}\right) }%
G_{k}(-1)\left( \nu (b)\right) ^{j_{2}+j_{3}+j_{4}+2}}{b^{3}\sqrt{\pi
^{3}B(j_{2})B(j_{3})B(j_{4})}}\frac{G_{k}(1+j_{2}+j_{3}+j_{4})}{G_{k}(-1)}%
\times  \notag \\
&&\times \frac{%
G_{k}(-j_{2}+j_{3}+j_{4})G_{k}(j_{2}-j_{3}+j_{4})G_{k}(j_{2}+j_{3}-j_{4})}{%
G_{k}(2j_{2}+1)G_{k}(2j_{3}+1)G_{k}(2j_{4}+1)}.  \label{nabajo}
\end{eqnarray}

It is instructive to compare (\ref{nabajo}) with (\ref{esaconelA}). This
realizes (\ref{ER}), and this relation between $X_{k}(0,j_{2},j_{3},j_{4})$
and $Y_{k^{\prime }}(0,j_{2},j_{3},j_{4})$ is somehow \textit{the inverse}
of that we found between (\ref{gato}) and (\ref{liebre}).

In the next subsection we will rederive relation (\ref{ER}) in a different
way. In particular, it will allow us to show how the Coulomb gas
representation emerges from the analytic extension of $%
Y_{k}(0,j_{2},j_{3},j_{4})$ to negative values of $k$ and $j_{i}$. In other
words, we will show that this relation between $SL(2,\mathbb{R})$ and $SU(2)$
WZNW models is nothing but the same sort of analytic continuation that one
considers in the free field representation of non-rational theories.

\subsection{The Coulomb gas approach and Wakimoto representation}

Here, we will reconsider the problem of how to recover $SL(2,\mathbb{R})_{k}$
structure constants from (\ref{Otraduda}). That is, we want to obtain%
\begin{equation*}
\left\langle \prod\nolimits_{i=2}^{4}\Phi _{j_{i}}(x_{i}|z_{i})\right\rangle
_{sl\text{(2)}}=\prod_{i<j}\left| x_{ij}\right|
^{2(j_{i}+j_{j}-j_{k})}\left| z_{ij}\right| ^{-2(\Delta _{i}+\Delta
_{j}-\Delta _{k})}C_{sl\text{(2)}}(j_{2},j_{3},j_{4}),
\end{equation*}%
with $C_{sl\text{(2)}}(j_{2},j_{3},j_{4})=X_{k}(0,j_{2},j_{3},j_{4}),$
starting from the expression for $Y_{k}(0,j_{2},j_{3},j_{4})$ in the $SU(2)$
case. So, let us consider the quantity 
\begin{equation*}
Y_{-k}(0,j_{2},j_{3},j_{4})\prod_{i=2}^{4}\sqrt{B(j_{i})}\sim \left( \nu
(b)\right) ^{s}\gamma (-b^{2})\prod_{i=2}^{4}\gamma \left(
1+b^{2}(2j_{i}+1)\right) \times
\end{equation*}

\begin{equation}
\times \frac{%
P_{-k}(s)P_{-k}(j_{2}+j_{3}-j_{4})P_{-k}(j_{2}-j_{3}+j_{4})P_{-k}(-j_{2}+j_{3}+j_{4})%
}{P_{-k}(2j_{2})P_{-k}(2j_{3})P_{-k}(2j_{4})}.  \label{xc1}
\end{equation}%
where $s=j_{2}+j_{3}+j_{4}+1$, and where the symbol $\sim $ stands for the
omission of irrelevant $b$-dependent factors. Let be also reminded of the
definition $P_{-k}(x)=\prod_{n=1}^{x}\gamma (-nb^{2})$ with $b^{-2}=k-2$.
Notice also that a divergent factor $\Gamma (0)$ arises in (\ref{xc1}),
although we are omitting it here. This factor stands for the integration
over the zero mode in the integral realization \cite{DiFK}, i.e. it
corresponds to the factor $\Gamma (-n)=\Gamma (-2j_{1})$ in (\ref{SSSS}),
(see (\ref{nuevaq})). This factor is eventually cancelled out by another
contribution $\Gamma ^{-1}(0)$ arising when analytically extending
expression (\ref{xc1}); see (\ref{xc5}) below.

The first step in rewriting (\ref{xc1}) will be to consider the three
factors of the form 
\begin{equation}
\frac{P_{-k}(j_{2}+j_{3}+j_{4}-2j_{a})}{P_{-k}(2j_{a})}=\frac{%
\prod_{r=1}^{j_{2}+j_{3}+j_{4}-2j_{a}}\gamma (-b^{2}r)}{\prod_{r=1}^{2j_{a}}%
\gamma (-b^{2}r)};
\end{equation}%
Let us write them by splitting the product. In turn, at least formally, we
can write 
\begin{equation*}
\frac{P_{-k}(j_{2}+j_{3}+j_{4}-2j_{a})}{P_{-k}(2j_{a})}=\prod_{r=1}^{2j_{a}}%
\gamma ^{-1}(-b^{2}r)\prod_{r=1}^{2j_{a}}\gamma
(-b^{2}r)\prod_{r=2j_{a}+1}^{j_{2}+j_{3}+j_{4}-2j_{a}}\gamma
(-b^{2}r)=\prod_{r=2j_{a}+1}^{j_{2}+j_{3}+j_{4}-2j_{a}}\gamma (-b^{2}r).
\end{equation*}%
Again, let us split the product, basically extending what would be valid for
the case $2j_{a}+1<-2j_{a}-1<j_{2}+j_{3}+j_{4}-2j_{a}$. Then, we write%
\begin{equation*}
\frac{P_{k}(j_{2}+j_{3}+j_{4}-2j_{a})}{P_{k}(2j_{a})}%
=\prod_{r=2j_{a}+1}^{-2j_{a}-1}\gamma
(-b^{2}r)\prod_{r=-2j_{a}}^{j_{2}+j_{3}+j_{4}-2j_{a}}\gamma (-b^{2}r).
\end{equation*}

Now, we can replace the products $\prod_{r=-x}^{x}\gamma (-b^{2}r)$
appearing in the expression above by the quantity $\left( -b^{2}\right)
^{-2x-1}\gamma (-x)$, using\footnote{%
It follows from prescription (\ref{Prescription}), but it can be also
heuristically motivated as follows: First consider the expansion $%
\prod_{r=-x}^{x}\gamma (-b^{2}r)=\frac{\Gamma (b^{2}x)\Gamma
(b^{2}(x-1))...\Gamma (b^{2})\Gamma (0)\Gamma (-b^{2})...\Gamma (-b^{2}x)}{%
\Gamma (1-b^{2}x)\Gamma (1-b^{2}(x-1))...\Gamma (1-b^{2})\Gamma (1)\Gamma
(1+b^{2})...\Gamma (1+b^{2}x)}$. Then, using $\Gamma (x+1)=x\Gamma (x)$ and
replacing $\Gamma (0)=(-1)^{-x}\Gamma (-x)\Gamma (x+1)$, one finds (\ref%
{DDDDDDD}).}%
\begin{equation}
\prod_{r=2j_{a}+1}^{-2j_{a}-1}\gamma (-b^{2}r)=(-b^{2})^{4j_{a}+1}\gamma
(2j_{a}+1)  \label{DDDDDDD}
\end{equation}

Then, (\ref{xc1}) would take the form%
\begin{eqnarray}
Y_{-k}(0,j_{2},j_{3},j_{4})\prod_{i=2}^{4}\sqrt{B(j_{i})} &\sim &\left( \nu
(b)\right) ^{s}\gamma (-b^{2})\left( -b^{2}\right) ^{4(1-s)}\prod_{a=2}^{4}%
\frac{\gamma \left( 2j_{a}+1\right) }{\gamma \left( -b^{2}(2j_{a}+1)\right) }%
\times  \notag \\
&&\times \prod_{r=1}^{s}\gamma
(-b^{2}r)\prod_{b=2}^{4}\prod_{r=-2j_{b}}^{j_{2}+j_{3}+j_{4}-j_{b}}\gamma
(-b^{2}r).  \label{haceletodo}
\end{eqnarray}

By manipulating $\Gamma $-functions, we get%
\begin{eqnarray}
Y_{-k}(0,j_{2},j_{3},j_{4})\prod_{i=2}^{4}\sqrt{B(j_{i})} &\sim &\left( \nu
(b)\right) ^{s}\gamma (-b^{2})\frac{\gamma (j_{2}-j_{3}-j_{4})}{\gamma
(2j_{2}+1)}\times  \notag \\
&&\times \frac{(-1)^{s}\mathcal{I}_{k}}{\pi ^{s}\gamma ^{s}(b^{2})\Gamma
(-s)\Gamma (1+s)}\prod_{a=2}^{4}\frac{\gamma \left( 2j_{a}+1\right) }{\gamma
\left( -b^{2}(2j_{a}+1)\right) }.  \label{trencitochucuchu}
\end{eqnarray}%
where we have defined

\begin{eqnarray}
\mathcal{I}_{k} &=&\Gamma (-s)\Gamma (s+1)\pi ^{s}(-1)^{s}\left(
-b^{2}\right) ^{2s}\left( \gamma (b^{2})\right) ^{s}\prod_{r=1}^{s}\gamma
(-b^{2}r)\times  \notag \\
&&\times \prod_{r=0}^{s-1}\left( \gamma (1-b^{2}(r-2j_{2}))\gamma
(-b^{2}(r-2j_{3}))\gamma (-b^{2}(r-2j_{4}))\right) .  \label{Acata}
\end{eqnarray}

The reason why we preferred to write the expression for $%
Y_{-k}(0,j_{2},j_{3},j_{4})\prod_{i=2}^{4}\sqrt{B(j_{i})}$ in its form (\ref%
{trencitochucuchu})-(\ref{Acata}) is that $\mathcal{I}_{k}$ can be
identified as the contribution coming from a Dotsenko-Fateev integral \cite%
{FZ,BB,GN3} 
\begin{equation}
\mathcal{I}_{k}=\Gamma (-s)\prod_{r=1}^{s}\int
d^{2}w_{r}\prod_{r=1}^{s}\left| w_{r}\right| ^{4j_{2}b^{2}}\left|
1-w_{r}\right| ^{4j_{3}b^{2}-2}\prod_{r<t}^{s-1,s}\left| w_{r}-w_{t}\right|
^{-4b^{2}}.  \label{xc3}
\end{equation}%
This follows from formula (B.9) of the Appendix of \cite{DF}.

It is worth noticing that integral (\ref{xc3}) is precisely the one that
arises in the Wakimoto free field representation of three-point functions 
\cite{BB}. For instance, the exponent of $\left| 1-w_{r}\right|
^{-2+4j_{3}b^{2}}$ in (\ref{xc3}) can be thought of as coming from the Wick
contraction between a $SL(2,\mathbb{R})_{k}$ vertex operator and the $r^{%
\text{th}}$ screening operator in the Coulomb gas representation. The
contributions $\left| w_{r}\right| ^{+4j_{2}b^{2}}$ indicate the presence of
highest weight states of discrete representations in the correlator.

Wakimoto free field representation follows from the considering the action%
\begin{equation}
S[\phi ,\beta ,\gamma ;\lambda ]=\frac{1}{4\pi }\int d^{2}z\left( \partial
\phi \overline{\partial }\phi -bR\phi /2\sqrt{2}+\overline{\beta }\partial 
\overline{\gamma }+\beta \overline{\partial }\gamma +4\pi \lambda \beta 
\overline{\beta }e^{-\sqrt{2}b\phi }\right) ,  \label{SWa}
\end{equation}%
where $\lambda $ is an arbitrary constant, $\beta (z)$ and $\gamma (z)$ form
a commutative ghost system, and $\phi (z)$ is a boson field with background
charge $-b=-1/\sqrt{k-2}$ \cite{Wakimoto}. The non-vanishing propagators are 
\begin{equation}
\left\langle \beta (w)\gamma (z)\right\rangle =\frac{1}{(w-z)},\qquad
\left\langle \phi (w)\phi (z)\right\rangle =-2\log |w-z|.
\end{equation}%
In the large $\phi $ regime, which corresponds to the near boundary limit in 
$AdS_{3}$ space, the vertex operators take the form%
\begin{equation*}
\Phi _{j_{i},m_{i},\overline{m}_{i}}(z_{i})=\gamma
_{(z_{i})}^{j_{i}+m_{i}}\gamma _{(\overline{z}_{i})}^{j_{i}+\overline{m}%
_{i}}e^{\sqrt{2}j_{i}b\phi (z_{i})}+B(j_{i})\gamma
_{(z_{i})}^{-1-j_{i}+m_{i}}\gamma _{(\overline{z}_{i})}^{-1-j_{i}+\overline{m%
}_{i}}e^{-\sqrt{2}(j_{i}+1)b\phi (z_{i})}+...
\end{equation*}%
with 
\begin{equation}
\Phi _{j_{i},m_{i},\overline{m}_{i}}(z_{i})=\int d^{2}x_{i}\Phi
_{j_{i}}(x_{i}|z_{i})x_{i}^{j_{i}+m_{i}}\overline{x}_{i}^{j_{i}+\overline{m}%
_{i}},
\end{equation}%
for $i=2,3,4$. On the other hand, the screening operators come from the
perturbation term in (\ref{SWa}), taking the form 
\begin{equation}
\mathcal{S}(w_{r})=\lambda \beta _{(w_{r})}\overline{\beta }_{(w_{r})}e^{-%
\sqrt{2}b\phi (w_{r})},
\end{equation}%
$r=1,2,...s$, with $s=j_{2}+j_{3}+j_{4}+1$.

This representation yields the integral expression (\ref{xc3}) through the
Wick contractions standing in%
\begin{equation*}
\lambda ^{s}\Gamma (-s)\prod\nolimits_{r=2}^{s}\int d^{2}w_{r}\left\langle
\prod\nolimits_{i=2}^{4}\gamma _{(z_{i})}^{j_{i}+m_{i}}\overline{\gamma }_{(%
\overline{z}_{i})}^{j_{i}+\overline{m}_{i}}e^{\sqrt{2}bj_{i}\phi
(z_{i})}\prod\nolimits_{r=2}^{s}\beta _{(w_{r})}\overline{\beta }_{(%
\overline{w}_{r})}e^{-\sqrt{2}b\phi (w_{r})}\right\rangle _{\lambda
=0}=\lambda ^{s}\mathcal{I}_{k}.
\end{equation*}%
where the average $\left\langle ...\right\rangle _{\lambda =0}$ is the
functional sum for the action (\ref{SWa}) with $\lambda =0$.

The precise relation between three-point functions in the $m$-basis and
those in the $x$-basis is discussed in \cite{GN3}. There, expressions like
the r.h.s. of (\ref{trencitochucuchu}) were shown to lead to exact result 
\cite{T1} through analytic continuation. Moreover, in \cite{GN3} (see Eqs.
(2.45) and (2.63) therein) it was discussed how the Dotsenko-Fateev integral
(\ref{xc3}) could be formally continued to be also expressed in terms of
special functions as follows 
\begin{eqnarray}
\mathcal{I}_{k} &=&b^{2}\pi ^{s}\left( \gamma (b^{2})\right) ^{s}\gamma
(-1-j_{2}-j_{3}-j_{4})\gamma (2j_{2}+1)\gamma (-j_{2}-j_{3}+j_{4})\gamma
(-j_{2}+j_{3}-j_{4})\times  \notag \\
&&\times \frac{G_{k}(-2-j_{2}-j_{3}-j_{4})}{G_{k}(-1)}\prod_{a=2}^{4}\frac{%
G_{k}(-1-j_{2}-j_{3}-j_{4}+2j_{a})}{G_{k}(-2j_{a}-1)},  \label{xc4}
\end{eqnarray}

The way of proposing expression (\ref{xc4}) is completely analog to what A.
Zamolodchikov and Al. Zamolodchikov did for LFT in \cite{ZZ}, where the
exact expression for Liouville structure constants was obtained from the
analytic continuation of the formula of the residues corresponding to
resonant correlators. Considering such analytic continuation, we can replace
the piece 
\begin{equation*}
\left( -b^{2}\right) ^{2s}\prod_{r=1}^{s}\gamma
(-b^{2}r)\prod_{r=0}^{s-1}\gamma (1-b^{2}(r-2j_{2}))\gamma
(-b^{2}(r-2j_{3}))\gamma (-b^{2}(r-2j_{4}))=
\end{equation*}%
\begin{equation*}
=\frac{(-1)^{s}}{\Gamma (-s)\Gamma (s+1)\pi ^{s}\gamma ^{s}(b^{2})}\mathcal{I%
}_{k},
\end{equation*}%
arising in (\ref{trencitochucuchu}), by the following contribution, 
\begin{eqnarray}
&&-\frac{\left( -b^{2}\right) ^{-2s+1}\gamma (-1-j_{2}-j_{3}-j_{4})\gamma
(-j_{2}-j_{3}+j_{4})\gamma (-j_{2}+j_{3}-j_{4})}{\Gamma (0)}\times  \notag \\
&&\times \frac{\gamma (2j_{2}+1)G_{k}(-2-j_{2}-j_{3}-j_{4})}{G_{k}(-1)}%
\prod_{a=2}^{4}\frac{G_{k}(-1-j_{2}-j_{3}-j_{4}+2j_{a})}{G_{k}(-2j_{a}-1)}.
\label{xc5}
\end{eqnarray}%
where the factor $\Gamma ^{-1}(0)$ arises from writing $(-1)^{-s}\Gamma
(-s)\Gamma (s+1)=\Gamma (0)$. As anticipated, this factor precisely cancels
the divergent factor $\Gamma (-2j_{1})=\Gamma (0)$ standing from evaluating $%
j_{1}=0$ in (\ref{SSSS}). Taking into account functional properties (\ref%
{Prop2}), one finds%
\begin{equation*}
Y_{-k}(0,j_{2},j_{3},j_{4})\prod_{i=2}^{4}\sqrt{B(j_{i})}\sim \left( \nu
(b)\right) ^{j_{2}+j_{3}+j_{4}+1}\frac{G_{k}(1+j_{2}+j_{3}+j_{4})}{G_{k}(-1)}%
\prod_{a=2}^{4}\frac{G_{k}(j_{2}+j_{3}+j_{4}-2j_{a})}{G_{k}(2j_{a}+1)}.
\end{equation*}

That is, we recovered $SL(2,\mathbb{R})_{k}$ structure constants from the
expression for $SU(2)_{k^{\prime }}$ model with $k^{\prime }=-k$; namely $%
Y_{-k}(0,j_{2},j_{3},j_{4})\prod_{i=2}^{4}\sqrt{B(j_{i})}\sim C_{sl\text{(2)}%
}(j_{2},j_{3},j_{4})$. This is nothing but (\ref{ER}), what we proved in the
previous subsection by means of the relation (\ref{kopada}).

\section{Discussion}

We have explained how the fact that three-point superstring amplitudes of
chiral states in $AdS_{3}\times S^{3}$ lead to a factorized expression does
not confront the fact that formulas of $SL(2,\mathbb{R})_{k}$ WZNW model can
be obtained from those of $SU(2)_{k^{\prime }}$ WZNW model by analytically
continuing in $k$. This turns out to be related to the shifting of the
Kac-Moody level $k$ in the supersymmetric theory: While in the bosonic
theory an appropriate analytic continuation of $SU(2)_{k^{\prime }}$
correlators leads to the expression of $SL(2,\mathbb{R})_{k}$ correlators
(with $k^{\prime }=-k$), in the supersymmetric theory both observables are,
roughly speaking, one the inverse of the other (with $k^{\prime }+2=k-2$).
In this sense, it is fair to say that the computation in the superstring
theory is more similar to the one in bosonic Minimal Liouville Gravity than
the one in bosonic WZNW model itself. It is the magic of supersymmetry what
is behind the cancellation in the three-point function, and not merely the
similarity between the Liouville theory and the $SL(2,\mathbb{R})$ WZNW
theory. This cancellation in the three-point function is the key point for
the matching between bulk and boundary observables \cite{GK,PD}, and this
was the motivation to revisit this calculation herein.

Before concluding, let us make some comments on the four-point function.
First, let us recall the relation between Liouville momenta $a_{i}$ and the
spin variable $j_{i}$ in the $SL(2,\mathbb{R})_{k}$ WZNW model ($k-2=b^{-2}$%
), namely%
\begin{equation*}
a_{1}=-\frac{b}{2}\left( j_{1}+j_{2}+j_{3}+j_{4}+1\right) ,\qquad a_{i}=-%
\frac{b}{2}\left( j_{1}+2j_{i}-j_{2}-j_{3}-j_{4}-b^{-2}-1\right) ,
\end{equation*}%
for $i=2,3,4$. On the other hand, the relation between the GMM momenta $%
\alpha _{i}$ and the $SU(2)_{k^{\prime }}$ WZNW model ($k^{\prime }+2=\beta
^{-2}$) spin variables $j_{i}^{\prime }$ is the following%
\begin{equation*}
\alpha _{1}=\frac{\beta }{2}\left( j_{1}^{\prime }+j_{2}^{\prime
}+j_{3}^{\prime }+j_{4}^{\prime }+1\right) ,\qquad \alpha _{i}=\frac{\beta }{%
2}\left( j_{1}^{\prime }+2j_{i}^{\prime }-j_{2}^{\prime }-j_{3}^{\prime
}-j_{4}^{\prime }+\beta ^{-2}-1\right) ,
\end{equation*}%
for $i=2,3,4$. Then, talking into account that in the supersymmetric theory $%
k^{\prime }+2=k-2$ and that chiral states obey $j_{i}=-1-j_{i}^{\prime }$,
we find%
\begin{equation}
a_{i}=\alpha _{i}+b\text{,}  \label{aba}
\end{equation}%
for the five states $i=1,2,3,4,5$. Remarkably, (\ref{aba}) is exactly the
relation between the momenta $\alpha _{i}$ and $a_{i}$ in Minimal Liouville
Gravity (MLG), as it is necessary for the vertex operators $V_{a_{i}}\times
W_{\alpha _{i}}$ to have conformal dimension one w.r.t the full
stress-tensor $T_{\text{Liouville}}+T_{\text{Minimal Model}}$. In turn,
restrictions on the momenta in the supersymmetric correlators in $%
AdS_{3}\times S^{3}\times T^{4}$ agree with requirements for conformal
invariance in the MLG.

Using (\ref{aba}) we can show that the expression for the bosonic part of
the worldsheet four-point functions $\left\langle \mathcal{O}_{j_{1}}%
\mathcal{O}_{j_{2}}\mathcal{O}_{j_{3}}\mathcal{O}_{j_{4}}\right\rangle $
simplifies in a remarkable way. Recalling%
\begin{equation*}
P_{k}(x)=\prod_{n=1}^{x}\gamma \left( n\beta ^{2}\right) =\frac{G_{k}(-1)}{%
G_{k}(-1-x)},\qquad x>0,
\end{equation*}%
and taking into account $j_{i}^{\prime }=-1-j_{i}$ (for $i=1,2,3,4$), we can
write the $SU(2)_{k^{\prime }}$ four-point function as follows%
\begin{eqnarray}
Y_{k^{\prime }}(-1-j_{1},-1-j_{2},-1-j_{3},-1-j_{4}) &=&\frac{\left( \gamma
(b^{2})\right) ^{-2j_{1}-1}}{G_{k}(2+\sum_{a=2}^{4}j_{a})}\prod_{n=1}^{4}%
\frac{G_{k}(2j_{n}+1)}{\sqrt{\gamma (-(2j_{n}+1)b^{2})}}\times  \notag \\
&&\times \frac{1}{\prod_{n=2}^{4}G_{k}(-1-2j_{n}-j_{1}+\sum_{i=2}^{4}j_{i})}.
\label{esaconelB}
\end{eqnarray}

Considering both (\ref{esaconelAA}) and (\ref{esaconelB}) together, the
final expression reads\footnote{%
Notice that there exists a remarkable similarity between this expression and
Eq. (\ref{RB}).} 
\begin{eqnarray*}
\mathcal{X}_{k}(j_{1},j_{2},j_{3},j_{4})\mathcal{Y}_{k^{\prime
}}(-1-j_{1},-1-j_{2},-1-j_{3},-1-j_{4}) &=&\frac{C_{W}^{2}}{\Upsilon _{0}^{2}%
}\frac{|z|^{2}|1-z|^{2}}{|x|^{2}|1-x|^{2}|z-x|^{2}}\times \\
&&\times \frac{\pi ^{3}}{b^{3+4b^{2}}}\prod\nolimits_{i=1}^{4}\frac{\sqrt{%
B(j_{i})}}{\gamma (2ba_{i}-b)}
\end{eqnarray*}%
where we have chosen $\mu \pi \gamma ^{2}(b^{2})b^{4-2b^{2}}=1$. Although
the computation of worldsheet four-point function, in addition, would
require to deal with the insertion of picture-changing operators in $%
\left\langle \mathcal{O}_{j_{1}}\mathcal{O}_{j_{2}}\widetilde{\mathcal{O}}%
_{j_{3}}\widetilde{\mathcal{O}}_{j_{4}}\right\rangle $, it is still
encouraging that the bosonic piece of the correlator $\left\langle \mathcal{O%
}_{j_{1}}\mathcal{O}_{j_{2}}\mathcal{O}_{j_{3}}\mathcal{O}%
_{j_{4}}\right\rangle $ yields a very simple form in terms of MLG five-point
functions. In fact, one gets%
\begin{equation*}
\left\langle \prod\nolimits_{i=1}^{4}\Phi _{j_{i}}(x_{i}|z_{i})\right\rangle
_{sl\text{(2)}}\times \left\langle \prod\nolimits_{i=1}^{4}\Psi
_{-1-j_{i}}(x_{i}|z_{i})\right\rangle _{su\text{(2)}}=\frac{\pi ^{3}}{%
b^{3+4b^{2}}}\frac{C_{W}^{2}}{\Upsilon _{0}^{2}}\prod\nolimits_{i=1}^{4}%
\frac{\sqrt{B(j_{i})}}{\gamma (2ba_{i}-b)}\times
\end{equation*}%
\begin{equation}
\times \frac{|z|^{2}|1-z|^{2}}{|x|^{2}|1-x|^{2}|z-x|^{2}}\left\langle
\prod\nolimits_{i=1}^{5}U_{a_{i}}(z_{i})\right\rangle _{\text{MLG}}
\label{carota}
\end{equation}%
where $\left\langle \prod_{i=1}^{5}U_{a_{i}}(z_{i})\right\rangle _{\text{MLG}%
}$ on the r.h.s. refers to the five-point correlation function in MLG; that
is%
\begin{equation}
\left\langle \prod\nolimits_{i=1}^{5}U_{a_{i}}(z_{i})\right\rangle _{\text{%
MLG}}=\left\langle \prod\nolimits_{i=1}^{5}V_{a_{i}}(z_{i})\right\rangle _{%
\text{LFT}}\times \left\langle
\prod\nolimits_{i=1}^{5}W_{a_{i}-b}(z_{i})\right\rangle _{\text{GMM}}
\end{equation}%
with $z_{2}=0$, $z_{3}=1$, $z_{4}=\infty $, while $z_{1}=z$, $z_{5}=x$. It
is worth mentioning that $N$-point correlation numbers in MLG were recently
computed \cite{FL,BZ,KP} for particular values of $N-3$ of the $N$ momenta $%
a_{i}$. Therefore, the fact one has access to these observables makes
relation (\ref{carota}) quite interesting. For instance, one could raise the
question whether holographic agreement for extremal four-point functions in $%
AdS_{3}\times S^{3}\times T^{4}$ is also observed as it happens in $%
AdS_{5}\times S^{5}$. To answer this kind of questions we have to learn more
about the non-renormalization mechanism and, more importantly, we have to
get more information about the boundary four-point function. Unfortunately,
four-point functions in the symmetric product to compare with are not
available; a computation of these observables would be a major progress.

\begin{equation*}
\end{equation*}

This work was partially supported by University of Buenos Aires, Agencia
ANPCyT, and CONICET, through grants UBACyT X861, PICT34557, and PIP6160.
G.G. thanks Ari Pakman and Leonardo Rastelli for previous collaboration and
for very interesting discussions. He is also grateful to Matt Kleban and the
members of the Center for Cosmology and Particle Physics CCPP of New York
University NYU for their hospitality during his stay, where this work was
finished.

\end{document}